\documentclass[11pt,a4paper]{article}
\usepackage{amsmath}
\usepackage{cite}
\usepackage{graphicx}
\usepackage{amsfonts}
\usepackage{amssymb}
\usepackage{psfig}

\def\bt{\beta}

\def\bt'{\beta'}

\begin{document}
\bigskip
\hfill\hbox{LPM-04-45} \vspace{2cm}

\begin{center}
{\Large \textbf{Form factors in the massless coset models\\
\vspace{0.5cm} \Large \textbf{$su(2)_{k+1}
\otimes su(2)_k /su(2)_{2k+1}$}}\\
\vspace{0.5cm}
\Large{\textbf{Part II}}}\\
\vspace{1.2cm} {\Large Paolo Grinza$^{a,}$ \footnote{
\textsf{grinza@lpm.univ-montp2.fr}} and B\'en\'edicte
Ponsot$^{b,}$ \footnote{\textsf
{benedicte.ponsot@anu.edu.au}}} \\
\vspace{0.7cm} {\it $^{a}$ Laboratoire de Physique Math\'ematique,
Universit\'e Montpellier II,\\
 Place Eug\`ene Bataillon, 34095
Montpellier Cedex 05, France}\\
\vspace{0.7cm} {\it $^{b}$Department of Theoretical Physics,\\
Research School of Physical Sciences and Engineering,\\
Australian National University,\\
Canberra, ACT 0200, Australia}
 \vspace{1cm}
\end{center}

\begin{abstract}
Massless flows from the coset model $su(2)_{k+1} \otimes su(2)_k
/su(2)_{2k+1}$ to the minimal model $M_{k+2}$ are studied from the
viewpoint of form factors. These flows include in particular the
flow from the Tricritical Ising model to the Ising model. By
analogy with the magnetization operator in the flow TIM $\to$ IM,
we construct all form factors of an operator that flows to
$\Phi_{1,2}$ in the IR. We make a numerical estimation of the
difference of conformal weights between the UV and the IR thanks
to the $\Delta$-sum rule; the results are consistent with the
conformal weight of the operator $\Phi_{2,2}$ in the UV. By
analogy with the energy operator in the flow TIM $\to$ IM, we
construct all form factors of an operator that flows to
$\Phi_{2,1}$. We propose to identify the operator in the UV with
$\sigma_1\Phi_{1,2}$.
\end{abstract}
\begin{center}
\end{center}

\newpage
\section*{Introduction}
In our previous article \cite{GP}, we considered the construction
of the form factors of the trace operator in the massless flows
\cite{CSS} from the UV coset model \cite{GKO} $su(2)_{k+1} \otimes
su(2)_k /su(2)_{2k+1}$, with central charge
$$
 c_{\textrm{\tiny UV}}=
\frac{3k(k+1)(2k+5)}{(k+2)(k+3)(2k+3)},
$$
 to the IR
coset $su(2)_{k} \otimes su(2)_1 /su(2)_{k+1}$. The latter model
is the unitary minimal model $M_{k+2}$ with central charge
$$
c_{\textrm{\tiny IR}}= \frac{k(k+5)}{(k+2)(k+3)}.
$$
The flow is defined in UV by the relevant operator of conformal
dimension $\Delta=\bar{\Delta}=1-2/(2k+3)$; it
arrives in the IR along the irrelevant operator $T\bar{T}$.\\
These flows include in particular for $k=1$ the famous massless
flow from the
 Tricritical Ising model to the Ising model. The latter flow was studied by Delfino, Mussardo and Simonetti
  in \cite{DMS} using a massless version of the form factor
  approach, originally developed for the massive case in \cite{KW,BKW,smirnov2}. For this purpose, the authors
   of \cite{DMS} used the scattering data proposed by Al.B.~Zamolodchikov in \cite{Z2}. Let us
   mention that the notion of
   massless scattering was first introduced and discussed in this
   latter paper.
In \cite{DMS}, beside the trace operator, some form factors for
the magnetization operator and the energy operator were
constructed.\\
 In this article we will try to generalize
this construction to the whole family of flows introduced above.\\
We do not intend to repeat here the construction of the form
factors in the Sine-Gordon model, and would like to refer the
reader to our previous paper \cite{GP} for notations and formulae, and to \cite{BFKZ,BK}
for complementary information on the global formalism.\\
Let us recall that in the massless case, the dispersion relations
read $(p^0,p^1)=\frac{M}{2}(e^{\theta},e^{\theta})$ for right
movers and $(p^0,p^1)=\frac{M}{2}(e^{-\theta'},-e^{-\theta'})$ for
left movers, where $M$ is some mass-scale in
 the theory, and $\theta,\theta'$ the rapidity variables. Zero momentum
  corresponds to $\theta\to -\infty$ for right movers and $\theta'\to +\infty$ for left movers.\\
The $S$-matrices for the three different scatterings were found in
\cite{Bernard}: the $RR$ and $LL$ $S$-matrices describe the IR CFT
$M_{k+2}$, and are thus given by the RSOS restriction of the
Sine-Gordon $S$-matrix \cite{L-RS,RS}.\\
We introduce by anticipation the minimal form factor in the SG
model:
\begin{eqnarray}
f_p(\theta)=-\mathrm{i}\sinh \frac{\theta}{2}
f^{min}_p(\theta)=-\mathrm{i}\sinh \frac{\theta}{2}\exp
\int_{0}^{\infty}\frac{dt}{t}\frac{\sinh\frac{1}{2}(1-p)t}{\sinh\frac{1}{2}p
t\cosh\frac{1}{2}t} \frac{1-\cosh
t(1-\frac{\theta}{\mathrm{i}\pi})}{2\sinh t}, \nonumber
\end{eqnarray}
where the parameter $p$ is related to the parameter $k$ by
$p\equiv k+2$. \\
As for the $RL$ scattering, it is given by\footnote{For the
particular cases $p=3$ and $p=+\infty$, the scattering data were
first proposed in \cite{Z2} and \cite{ZZ}
respectively.}\cite{Bernard}:
\begin{eqnarray}
S_{RL}(\theta-\theta')=\frac{1}{S_{LR}(\theta'-\theta)}=\tanh\left(\frac{\theta-\theta'}{2}-\frac{\mathrm{i}\pi}{4}\right)
\, . \nonumber
\end{eqnarray}
In the IR limit, the scattering becomes trivial $S_{RL}(\theta-\theta') \to -1$, and the two chiralities decouple.\\
The
minimal form-factor in the RL channel satisfies the relation:
$$
f_{RL}(\theta-\theta')=f_{RL}(\theta-\theta'+2\mathrm{i}\pi)S_{RL}(\theta-\theta'),
$$
and its explicit expression is given by
\begin{eqnarray}
f_{RL}(\theta-\theta')=\exp\left(
\frac{(\theta-\theta'-\mathrm{i}\pi)}{4}-\int_{0}^{+\infty}\;
 \frac{dt}{t}\frac{1-\cosh t\left(1-\frac{(\theta-\theta')}{\mathrm{i}\pi}\right)}
 {2\sinh t\cosh \frac{t}{2}}\right).
\nonumber
\end{eqnarray}
Its asymptotic behaviour in the infrared is:
$f_{RL}(\theta-\theta')\sim \mathcal{K}\;
e^{\frac{1}{2}(\theta-\theta'-\mathrm{i}\pi)}$, where
\begin{eqnarray}
\mathcal{K}=\exp{-\frac{1}{2}\int_{0}^{\infty}\frac{dt}{t}\left(\frac{1}{\sinh
t \cosh \frac{t}{2}}-\frac{1}{t}\right)}. \nonumber
\end{eqnarray}
\vskip0.4cm The plan of the paper is the following: in Section 1,
we generalize the construction of the form factors of the
magnetization operator in the flow TIM $\to$ IM, {\it i.e.} we
construct form factors of an operator that flows to $\Phi_{1,2}$
in the IR. Then we make numerical checks involving the
$\Delta$-sum rule, in order to identify the conformal dimension of
the operator in the UV. Section 2 is devoted to the generalization
of the form factors of the energy operator in the flow TIM $\to$
IM: we construct form factors of an operator which flows in the IR
to the operator $\Phi_{2,1}$ in $M_{k+2}$; an analysis of the two
point correlator for such an operator can also be found. Finally,
we give some concluding remarks.

\section{Form factors of an operator $\Phi$ that flows to $\Phi_{1,2}$
in $M_{p}$.}
\subsection{Magnetization operator in the flow
from TIM to IM.} A few form factors of the magnetization operator
 were obtained in \cite{DMS} in terms of symmetric polynomials
for
 the case $k=1$ ($p=3$) corresponding to the massless flow from TIM to IM. Due to the invariance of the theory
 under spin reversal,
 the order operator
 has non vanishing form factors only on a odd number of particles, whereas the disorder operator only on an even one.
 We shall consider the case of the disorder operator with even number of right and left
 particles.\\
We recall
  that the form factors of this operator satisfy the following residue equation
  at $\theta_1=\theta_{2r}+\mathrm{i}\pi$ \cite{DMS,YZ}:
\begin{eqnarray}
\mathrm{res}F_{2r,2l}(\theta_1,\cdots,\theta_{2r};\theta'_1,\cdots,\theta'_{2l})=
-2\mathrm{i}\;F_{2r-2,2l}(\theta_2,\cdots,\theta_{2r-1};\theta'_1,\cdots,\theta'_{2l})
\left(1+\prod_{k=1}^{2l}S_{RL}(\theta_{2r}-\theta'_k)\right),\nonumber
\end{eqnarray}
and a similar equation in the $LL$ channel. The first step of recursion is given by $F_{0,0}=1$.\\
In \cite{P} it was observed that all form factors of this operator
could be rewritten as:
\begin{eqnarray}
\lefteqn{F_{2r,2l}(\theta_1,\ldots,\theta_{2r};\theta'_1,\ldots,\theta'_{2l})=}
\nonumber \\
&& \prod_{1\leq i<j\leq 2r}\sinh\frac{\theta_{ij}}{2}\prod_{1\leq
i<j\leq 2l}\sinh\frac{\theta'_{ij}}{2} \prod_{i,j}f_{RL}(\theta_i
-\theta'_j)
Q_{2r,2l}(\theta_1,\ldots,\theta_{2r};\theta'_1,\ldots,\theta'_{2l}),
\nonumber
\end{eqnarray}
where
\begin{eqnarray}
    \lefteqn{Q_{2r,2l}(\theta_1,\ldots,
      \theta_{2r};\theta'_1,\ldots,\theta'_{2l})=}\nonumber \\
    && (-\mathrm{i}\pi)^{r+l}\sum_{T \subset S, \atop \#T=r}\sum_{T' \subset S', \atop
      \#T'=l}\prod_{i \in T, \atop k\in
      \bar{T}}\phi(\theta_{ik})e^{\frac{1}{2}\sum\theta_{ki}}\prod_{i
      \in T',\atop  k\in \bar{T}'}\phi(\theta'_{ik})
    e^{\frac{1}{2}\sum\theta'_{ik}} \prod_{i \in T, \atop k\in
      \bar{T}'}\Phi(\theta_i-\theta'_k) \prod_{i \in T', \atop k\in
      \bar{T}}\tilde{\Phi}(\theta_k-\theta'_i)\nonumber,
  \end{eqnarray}
  and
  \begin{eqnarray}
\phi(\theta_{ij}) \equiv
\frac{-1}{f_{RR}(\theta_{ij})f_{RR}(\theta_{ij}+\mathrm{i}\pi)}
=\frac{2\mathrm{i}}{\sinh \theta_{ij}}\; ,\quad \phi(\theta'_{ij})
\equiv
\frac{-1}{f_{LL}(\theta'_{ij})f_{LL}(\theta'_{ij}+\mathrm{i}\pi)}
= \frac{2\mathrm{i}}{\sinh \theta'_{ij}}\; ,\label{phi}
\end{eqnarray}
as well as:
\begin{eqnarray}
\Phi(\theta-\theta')\equiv
\frac{S_{RL}(\theta-\theta')}{f_{RL}(\theta-\theta')f_{RL}(\theta-\theta'+\mathrm{i}\pi)}=
 \mathcal{K}^{-2}(1-\mathrm{i}e^{\theta'-\theta}),\quad
\tilde{\Phi}(\theta-\theta') \equiv
\Phi(\theta-\theta'+\mathrm{i}\pi). \label{grandphi}
\end{eqnarray}
It was found in \cite{DMS} that for two right and two left
particles, the expression of the form factor is given by the
following expression:
\begin{eqnarray}
F_{2,2}(\theta_1,\theta_2;\theta'_1,\theta'_2)=\frac{1}{4\mathcal{K}^{4}}
\tanh\frac{\theta_{12}}{2}\tanh\frac{\theta'_{12}}{2} \prod_{i=1,2
\atop j=1,2}f_{RL}(\theta_i-\theta'_j)\;
(1+e^{\theta'_1+\theta'_2-\theta_1-\theta_2}). \label{disorder}
\end{eqnarray}

 The following
relation holds when $\theta_i-\theta'_j \to -\infty$ which defines
the IR region \cite{DMS,P}:
\begin{eqnarray}
F_{2r,2l}(\theta_1,\ldots,\theta_{2r};\theta'_1,\ldots,\theta'_{2l})\to
F_{2r}^{\mu}(\theta_1,\ldots,\theta_{2r})
F_{2l}^{\mu}(\theta'_1,\ldots,\theta'_{2l}), \nonumber
\end{eqnarray}
where $F_{2r}^{\mu}(\theta_1,\ldots,\theta_{2r})$ are the form
factors of the disorder operator $\mu=\Phi_{1,2}$ in the thermal
Ising model \cite{BKW,YZ,CM}:
$$
F_{2r}^{\mu}(\theta_1,\ldots,\theta_{2r})=\frac{1}{2^{r(2r-1)}}\prod_{1\leq
i<j<\leq 2r} \tanh\frac{\theta_{ij}}{2}.
$$

\subsection{Generalization} By analogy with the previous section,
we shall now look for a solution to the following problem at
$\theta_1=\theta_{2r}+\mathrm{i}\pi$:
\begin{eqnarray}
\lefteqn{\mathrm{res}F_{2r,2l}^{\Phi}(\theta_1,\cdots,\theta_{2r};\theta'_1,\cdots,\theta'_{2l})=}\label{RES} \\
&&
-2\mathrm{i}\;F_{2r-2,2l}^{\Phi}(\theta_2,\cdots,\theta_{2r-1};\theta'_1,\cdots,\theta'_{2l})
\left(1+\prod_{i=2}^{2r-1}S^{RSOS}_p(\theta_{i}-\theta_{2r})
\prod_{k=1}^{2l}S_{RL}(\theta_{2r}-\theta'_k)\right)
e^{R},\nonumber
\end{eqnarray}
where $e^{R}=e^{\frac{\mathrm{i}\pi}{2p}}s_1\otimes
\bar{s}_{2r}+e^{-\frac{\mathrm{i}\pi}{2p}}\bar{s}_1\otimes
s_{2r}$. A similar equation holds in the $LL$ channel.\\
Let us note that in the IR limit, given that $S_{RL} \to -1$, the
latter relation becomes
\begin{eqnarray}
\lefteqn{\mathrm{res}F_{2r,2l}^{\Phi}(\theta_1,\cdots,\theta_{2r};\theta'_1,\cdots,\theta'_{2l})=}\nonumber \\
&&
-2\mathrm{i}\;F_{2r-2,2l}^{\Phi}(\theta_2,\cdots,\theta_{2r-1};\theta'_1,\cdots,\theta'_{2l})
\left(1+\prod_{i=2}^{2r-1}S^{RSOS}_p(\theta_{i}-\theta_{2r})\right)
e^{R},\nonumber
\end{eqnarray}
which is the residue equation satisfied by (amongst others) the
operator $\Phi_{1,2}$ in $M_{p}$.\\
 We construct a solution to the residue
equation (\ref{RES}) with the initial condition
$F_{0,0}^{\Phi}=\langle\Phi\rangle$, and with the following
condition in the IR limit:
\begin{eqnarray}
F_{2r,2l}^{\Phi}(\theta_1,\cdots,\theta_{2r};\theta'_1,\cdots,\theta'_{2l})
\to \langle\Phi\rangle \;
f_{RSOS}^{\Phi_{1,2}}(\theta_1,\cdots,\theta_{2r})\;
f_{RSOS}^{\Phi_{1,2}}(\theta'_1,\cdots,\theta'_{2l}). \label{IR}
\end{eqnarray}
In other words, we want to construct form factors of an operator
that renormalizes in the IR on the operator $\Phi_{1,2}$ in
$M_{p}$. \\
We make the following ansatz for the solution, to be compared with
the one obtained for the trace operator in \cite{GP} (once again,
we refer the reader to \cite{GP,BFKZ,BK} for further explanations
on the construction of form factors in the SG model and basic
notations):
\begin{eqnarray}
\lefteqn{F_{2r,2l}^{\Phi}(\theta_1,\dots,\theta_{2r};\theta'_{1},\dots,\theta'_{2l})=
\langle\Phi\rangle N^{\Phi}_{2r}N^{\Phi}_{2l}\prod_{1\le i<j\le
2r}f_p(\theta_{ij})\prod_{1\le i<j\le 2l}f_p(\theta'_{ij})
\prod_{i,j}f_{RL}(\theta_i-\theta'_{j})
} \nonumber \\
&& \times \int_{C_{\theta}}\prod_{m=1}^r du_m\;
h_{RR}(\theta,u)p^{\frac{1}{2}}_{2r}(\theta,u)
\tilde{\Psi}^{p}(\theta,{u})\; \int_{C_{\theta'}}\prod_{m=1}^l
dv_m\; h_{LL}(\theta',v)
p^{\frac{1}{2}}_{2l}(\theta',v) \tilde{\Psi}^{p}(\theta',{v})\nonumber \\
&& \times \; \mathcal{N}_{2r,2l}(\theta,\theta',{u},{v}).
\label{FPhi}
\end{eqnarray}
We introduced $p^{\frac{1}{2}}(\theta,u)$, which is the
$p$-function\footnote{It is the only ingredient in the formula
above that depends explicitly on the operator considered, see
\cite{BK}.} of the operator $\Phi_{1,2}$ in the minimal model
$M_{p}$. We use the identification $\Phi_{1,2} \sim
e^{\mathrm{i}\frac{\beta}{2} \varphi_{SG}}$, followed by a
modification of the multiparticles state \cite{RS}: the modified
Bethe ansatz state $\tilde{\Psi}^{p}$ is related to the usual
Bethe ansatz state by the relation \cite{RS}
\begin{eqnarray}
\tilde{\Psi}^{p}_{\epsilon_1 \epsilon_2 \dots
\epsilon_{2n}}(\theta,u)\equiv e^{\frac{1}{2p}\sum_i \epsilon_i
\theta_i} \Psi^{p}_{\epsilon_1 \epsilon_2 \dots
\epsilon_{2n}}(\theta,u), \quad \epsilon_i=\pm,\;
\sum_{i=1}^{2n}\epsilon_i=0. \nonumber
\end{eqnarray}
We will use the $p$-function of the exponential fields
$e^{\mathrm{i}\alpha\varphi}$ in the SG model for the particular
value $\alpha=\beta/2$ (the form factors of
$e^{\mathrm{i}\alpha\varphi}$ were first constructed in \cite{L};
we use here the conventions and notations of \cite{P2}):
\begin{eqnarray}
p^{\frac{1}{2}}_{2n}(\theta,u)=\frac{1}{e^{\frac{\mathrm{i}\pi
}{2}}} \frac{\prod_{j=1}^{n} e^{u_j}}{\prod_{i=1}^{2n}
e^{\frac{\theta_i}{2}}}. \nonumber
\end{eqnarray}
We introduced the scalar function (completely determined by the
$S$-matrix)
\begin{eqnarray}
h_p(\theta,u)=\prod_{i=1}^{2n}\prod_{j=1}^{m}\phi_p(\theta_{i}-u_j)
\prod_{1\le r<s\le m}\tau_p(u_r-u_s),\nonumber
\end{eqnarray}
with
$$
\phi_p(u)=\frac{1}{f_p(u)f_p(u+\mathrm{i}\pi)},\quad
\tau_p(u)=\frac{1}{\phi_p(u)\phi_p(-u)},
$$
and $N^{\Phi}_{2r},N^{\Phi}_{2l}$ are normalization constants.
\\
Finally, the integration contours $C_{\theta}$ consist of several
pieces for all integration variables $u_j$~: a line from $-\infty$
to $\infty$
 avoiding all poles such that
$\mathrm{Im}\theta_i-\pi-\epsilon<\mathrm{Im}
u_j<\mathrm{Im}\theta_i-\pi,$ and clockwise oriented circles
around the poles (of the $\phi(\theta_i-u_j)$) at $\theta_i=u_j$,
$(j=1,\dots,m)$. The integration contours $C_{\theta'}$ are similarly defined.
The function $\mathcal{N}_{2r,2l}$ remains to be determined.\\
Let us motivate our ansatz (\ref{FPhi}): it is a slight
modification of the results of \cite{BFKZ, BK} that the form
factors of the operator $\Phi_{1,2}$ in the minimal model $M_{p}$
are written:
\begin{eqnarray}
f_{RSOS}^{\Phi_{1,2}}(\theta_1,\cdots,\theta_{2r})=N^{\Phi}_{2r}\prod_{1\le
i<j\le 2r}f_p(\theta_{ij})\int_{C_{\theta}}\prod_{m=1}^r du_m\;
h_{RR}(\theta,u)p^{\frac{1}{2}}_{2r}(\theta,u)
\tilde{\Psi}^{p}(\theta,{u}).\nonumber
\end{eqnarray}
When $\theta_{1}=\theta_{2r}+\mathrm{i}\pi$, each of the $r$
integration contours gets
 pinched at $\theta_{2r},\theta_{2r}\pm \mathrm{i}\pi$, and we
 have to take the sum of these three contributions.
 Due to symmetry, it is enough to consider the contribution of one of them ({\it e.g.} $u_r$), and multiply the result by
 $r$. The following computation is detailed in \cite{BFKZ}:\\
 at $u_r=\theta_{2r}$, we have:
\begin{eqnarray}
\mathrm{res}f_{RSOS}^{\Phi_{1,2}}(\theta_1,\cdots,\theta_{2r})=
-2\mathrm{i}\;f_{RSOS}^{\Phi_{1,2}}(\theta_2,\cdots,\theta_{2r-1})e^{R},\nonumber
\end{eqnarray}
whereas at $u_r=\theta_{2r}+\mathrm{i}\pi$:
\begin{eqnarray}
\lefteqn{\mathrm{res}f_{RSOS}^{\Phi_{1,2}}(\theta_1,\cdots,\theta_{2r})=}\nonumber \\
&&
-2\mathrm{i}\;f_{RSOS}^{\Phi_{1,2}}(\theta_2,\cdots,\theta_{2r-1})
\prod_{i=2}^{2r-1}S^{RSOS}_p(\theta_{i}-\theta_{2r})(e^{\frac{\mathrm{i}\pi}{2p}}s_1\otimes
\bar{s}_{2r}),\nonumber
\end{eqnarray}
and at $u_r=\theta_{2r}-\mathrm{i}\pi$:
\begin{eqnarray}
\lefteqn{\mathrm{res}f_{RSOS}^{\Phi_{1,2}}(\theta_1,\cdots,\theta_{2r})=}\nonumber \\
&&
-2\mathrm{i}\;f_{RSOS}^{\Phi_{1,2}}(\theta_2,\cdots,\theta_{2r-1})
\prod_{i=2}^{2r-1}S^{RSOS}_p(\theta_{i}-\theta_{2r})(e^{-\frac{\mathrm{i}\pi}{2p}}\bar{s}_1\otimes
s_{2r}).\nonumber
\end{eqnarray}
Remembering the relation:
\begin{eqnarray}
f_{RL}(\theta-\theta')f_{RL}(\theta-\theta'+\mathrm{i}\pi)=\frac{\mathcal{K}^2}{1+\mathrm{i}e^{\theta'-\theta}},\nonumber
\end{eqnarray}
we see that the ansatz (\ref{FPhi}) will satisfy the residue
equation (\ref{RES}) at the condition that the function
$\mathcal{N}_{2r,2l}(\theta,\theta',{u},{v})$ satisfies at
$\theta_{1}=\theta_{2r}+\mathrm{i}\pi$:
\begin{itemize}
\item
 $u_r=\theta_{2r}$:
\begin{eqnarray}
\lefteqn{\mathcal{N}_{2r,2l}(\theta_1,\dots,\theta_{2r};\theta';u_1,\dots
,u_r;v)=}
\nonumber \\
&& \mathcal{N}_{2r-2,2l} (\theta_2,\dots,\theta_{2r-1};\theta';
u_1,\dots, u_{r-1};v)
 \prod_{k=1}^{2l}\tilde\Phi(\theta_{2r}-\theta'_k).
 \label{rel1}
\end{eqnarray}
\item
 $u_r=\theta_{2r}\pm \mathrm{i}\pi$:
\begin{eqnarray}
\lefteqn{\mathcal{N}_{2r,2l}(\theta_1 \dots \theta_{2r};\theta';
u_1 \dots u_r;v)=}\nonumber \\
&& \mathcal{N}_{2r-2,2l} (\theta_2,\dots,\theta_{2r-1};\theta';
u_1,\dots, u_{r-1};v)
\prod_{k=1}^{2l}{\Phi}(\theta_{2r}-\theta'_k), \label{rel2}
\end{eqnarray}
and similar relations in the $LL$ channel. \end{itemize}
We
introduce the sets $S=(1,\dots,2r)$ and $S'=(1,\dots,2l)$, as well
as $T,U,V$ the subsets of $S$, such that
$$
S=T\cup U\cup V,
$$
with $T\cap U=\emptyset$ and $V=S-(T\cup V)$. These subsets have
number of elements: $\#T=r-1$, $\#U=1$, $\#V=r$.
\begin{eqnarray}
T = \{i_1<i_2<\dots<i_{r-1}\},\quad U=\{i_u\},\quad V =
\{k_1<k_2<\dots<k_{r}\}. \nonumber
\end{eqnarray}
The subsets of $S'$: $T',U'$ and $V'$ are defined similarly.\\
We conjecture
 the following expression:
\begin{eqnarray}
\lefteqn{\mathcal{N}_{2r,2l}(\theta,\theta',{u},{v})=\frac{1}{\sum_{i=1}^{2r}
 e^{\theta_i}\sum_{i=1}^{2l} e^{-\theta'_i}} \times \quad \sum_{T\subset S}\sum_{U\subset S}\sum_{T'\subset
S'}\sum_{U'\subset S' }}\nonumber \\
&& \times \; \frac{\prod_{k,l\in V\atop k<l
}\cos\frac{\theta_{kl}}{2\mathrm{i}} \prod_{k\in
V}\cos\frac{\theta_{ku}}{2\mathrm{i}} \prod_{i\in
T}\cos\frac{\theta_{iu}}{2\mathrm{i}}} {\prod_{i\in T,\atop k\in
V}\sin\frac{\theta_{ik}}{2\mathrm{i}} \prod_{i\in
T}\sin\frac{\theta_{ui}}{2\mathrm{i}}
 \prod_{k\in V}\sin\frac{\theta_{ku}}{2\mathrm{i}}}e^{\theta_u}\;\frac{\prod_{k,l\in V'\atop
k<l}\cos\frac{\theta'_{kl}}{2\mathrm{i}}\prod_{k\in
V'}\cos\frac{\theta'_{ku}}{2\mathrm{i}} \prod_{i\in
T'}\cos\frac{\theta'_{iu}}{2\mathrm{i}}  }{\prod_{i\in T',\atop
k\in V'} \sin\frac{\theta'_{ik}}{2\mathrm{i}}\prod_{i\in T'}
\sin\frac{\theta'_{ui}}{2\mathrm{i}} \prod_{k\in V'}
\sin\frac{\theta'_{ku}}{2\mathrm{i}} }e^{-\theta'_u}\nonumber\\
&&  \times \;
 \frac{\prod_{i\in T,\atop m=1,\dots
,r}\cos\frac{\theta_i-u_m}{2\mathrm{i}} \prod_{i\in T',\atop
m=1,\dots ,l}\cos\frac{\theta'_i-v_m}{2\mathrm{i}}}{\prod_{1\le
m<n\le r} \cos \frac{u_m-u_n}{2\mathrm{i}}\prod_{1\le m<n\le
l}\cos \frac{v_m-v_n}{2\mathrm{i}}} \prod_{i \in T\cup U, \atop
k\in V'}\tilde\Phi(\theta_i-\theta'_k) \prod_{i \in T'\cup U',
\atop k\in V}\Phi(\theta_k-\theta'_i)\; \nonumber .
\end{eqnarray}
Let us give a sketch of the proof that this function satisfies the
relations (\ref{rel1}) and (\ref{rel2}):
\begin{itemize}
\item
when $\theta_1=\theta_{2r}+\mathrm{i}\pi$ and $u_r=\theta_{2r}$,
we cannot have $\theta_1 \in T$, otherwise the term $\cos
\frac{\theta_1-u_r}{2\mathrm{i}}$ becomes equal to zero.
Consequently we should have $\theta_1 \in U$ or $\theta_1 \in V$.
It follows from a simple inspection of the other cosines in the
numerator that the only possibility in order to have the function
$\mathcal{N}$ different from zero is to have $\theta_1 \in V$ and
$\theta_{2r} \in T$. Equation (\ref{rel1}) follows from the use of
simple trigonometric identities.
\item
conversely, when $\theta_1=\theta_{2r}+\mathrm{i}\pi$ and
$u_r=\theta_{2r}\pm \mathrm{i}\pi$, we find that we should have
$\theta_1 \in T$ and $\theta_{2r} \in V$. Equation (\ref{rel2})
follows.
\end{itemize}
Moreover, it was checked with Mathematica (for a small number of
particles) that the relation (\ref{IR}) holds.

\subsection{Numerical results}
By analogy with what was done in \cite {DSC} for the flow TIM
$\to$ IM, one can think of using the $\Delta$-sum rule in
order to compute the variation of the conformal dimension and
hence identify the conformal dimension of the operator in the UV.
We recall the relation \cite{DSC}:
\begin{eqnarray}
D\equiv \Delta^{\textrm{\tiny UV}}_{\Phi}-\Delta^{\textrm{\tiny
IR}}_{\Phi}=-\frac{1}{2 \langle \Phi \rangle}\int_{0}^{\infty}dr\;
r \;\langle \Theta(r)\Phi(0) \rangle, \label{delta}
\end{eqnarray}
where $\langle\Phi\rangle$ is the vacuum expectation value of the
operator
$\Phi$, and $\Theta$ the trace operator. \\
 For our purpose, we will need the expression of the
4-particles form factor of $\Theta$ found in \cite{GP}:
\begin{eqnarray}
\lefteqn{F^{\Theta}(\theta_1,\theta_2;\theta'_1,\theta'_2) }\nonumber\\
&& = \frac{16\pi M^2 }{p^2\mathcal{K}^4}\;
f_p(\theta_{12})f_p(\theta'_{12})\prod_{i=1,2 \atop
j=1,2}f_{RL}(\theta_{i}-\theta'_{j})\;
\frac{\cosh\frac{\theta_{12}}{2}\cosh\frac{\theta'_{12}}{2}\;
e^{R}\otimes
e^L}{\sinh\frac{1}{p}(\mathrm{i}\pi-\theta_{12})\sinh\frac{1}{p}(\mathrm{i}\pi-\theta'_{12})},\nonumber
\end{eqnarray}
as well as the 4-particles form factor of the operator $\Phi$. The
formula presented in the previous section considerably simplifies
for 4 particles:
\begin{eqnarray}
\mathcal{N}_{2,2}(\theta_1,\theta_2;\theta'_1,\theta'_2;u,v)=\mathcal{K}^{-4}(1+e^{\theta'_1+\theta'_2-\theta_1-\theta_2}),\nonumber
\end{eqnarray}
and this leads to the following expression for the form factor:
\begin{eqnarray}
\lefteqn{F^{\Phi}(\theta_1,\theta_2;\theta'_1,\theta'_2)}\nonumber
\\
&&
=\frac{\langle\Phi\rangle}{p^2\mathcal{K}^{4}}f_{p}(\theta_{12})f_{p}(\theta'_{12})
\prod_{i=1,2 \atop j=1,2}f_{RL}(\theta_i-\theta'_j) \;
\frac{(1+e^{\theta'_1+\theta'_2-\theta_1-\theta_2})\;e^{R}\otimes
e^L}{\sinh \frac{1}{p}(\mathrm{i}\pi-\theta_{12})\sinh
\frac{1}{p}(\mathrm{i}\pi -\theta'_{12})}.\label{FPhi1}
\end{eqnarray}
The normalization of the latter form factor was chosen in order to
ensure the initial condition $F^{\Phi}_{0,0}=\langle\Phi\rangle$.
This amounts to setting the constant:
$N^{\Phi}_2=\frac{\mathcal{C}^4_p}{\pi p}$ in formula
(\ref{FPhi}), where $\mathcal{C}^4_p$ is defined by the asymptotic
behaviour when $\theta \to \pm \infty$ of the minimal form factor:
$$
f_p(\theta)\sim \mathcal{C}_p\;
e^{\pm\frac{1}{4}(\frac{1}{p}+1)(\theta-\mathrm{i}\pi)}.
$$
Explicitly:
\begin{eqnarray}
\mathcal{C}_p=\frac{1}{2}\exp
\frac{1}{2}\int_{0}^{\infty}\frac{dt}{t}
\left(\frac{\sinh\frac{1}{2}(1-p)t}{\sinh\frac{1}{2}p
t\cosh\frac{1}{2}t\sinh t}-\frac{1-p}{p t}\right).\nonumber
\end{eqnarray}
The four particle form factor (\ref{FPhi1}) at $p=3$ should be
compared with equation (\ref{disorder}); in this case we have:
\begin{eqnarray}
\frac{f_{3}(\theta_{12})}{\sinh
\frac{1}{3}(\mathrm{i}\pi-\theta_{12})}=-\frac{3}{2}\tanh\frac{\theta_{12}}{2}\nonumber,
\end{eqnarray}
and consequently the two expressions are identical. \\
For an arbitrary number of particles, our general construction
(\ref{FPhi}) for $p=3$ should reduce to
  the formula presented in section 1.1. This looks quite a non trivial
  task to be performed,
   and we hope we can return to this issue in the future
(a similar situation has already occured in \cite{GP} and \cite{P2}).\\
 Let us note finally that the v.e.v
$\langle\Phi\rangle$ does not need
to be known exactly, as it does not enter the numerical estimation.\\
In order to apply the $\Delta$-sum rule test, it is important to
have in mind a UV operator that could be a good candidate: we
recall that the numerical tests on the central charge in \cite{GP}
already showed quite a large discrepancy with respect to the exact
results, so we do not expect the sum rule for the
conformal dimension to give particularly accurate results. \\
The minimal model $M_{k+2}$ is nothing but the coset model
$su(2)_{k} \otimes su(2)_1 /su(2)_{k+1}$, and we will denote the
latter $\mathcal{M}(k+2,3)$. The operator $\Phi_{1,2}$ in
$\mathcal{M}(k+2,3)$ has conformal dimension
$$\Delta_{1,2}^{\textrm{\tiny{IR}}}=\frac{k}{4(k+3)},
$$ which is the same as $\Phi_{2,2}$ in the coset model\footnote{In the coset
 model $\mathcal{M}(k+2,l+2)$, the primaries $\sigma_{j}\Phi_{r,s}$ with $0<r<k+2,\; 0<s<k+l+2$ have conformal weight $\Delta=
 \frac{((k+2)s-(k+l+2)r)^2-l^2}{4l(k+2)(k+l+2)} +\frac{j(l-j)}{2l(l+2)}$, where $j=0,1,\dots,l/2 \;[(l-1)/2]$ for $l$ even [odd]
 and $|s-r|=|j\;\textrm{mod}\; l|$.} $\mathcal{M}(3,k+2)$. From
the perturbative RG calculations \cite{LC,Z4} for the $\Phi_{1,3}$
induced flow from $M_{k+3}$ to $M_{k+2}$,
 and the Landau-Ginzburg representation \cite{LC,Z5}, it is well known that one expects $\Phi_{2,2}$ in $M_{k+3}$
  to flow to $\Phi_{2,2}$ in $M_{k+2}$ for $k$ odd. For $k=1$, $\Phi_{2,2}$
  in $M_{4}$ has conformal dimension $\frac{3}{80}$,
   and $\Phi_{2,2}=\Phi_{1,2}$ in $M_{3}$
has conformal dimension $\frac{1}{16}$. This flow was later
confirmed by \cite{KM} by guessing
 massless TBA equations for the first excited state in the theory $M_{4}$, via a suitable
 modification of
  the massless TBA equations for the ground state found in
  \cite{Z2}.\\
   Then, using the form factors obtained in \cite{DMS} for
  the flow TIM $\to$ IM, the authors of \cite{DSC} successfully
  identified the conformal dimension of $\Phi_{2,2}$ in the UV thanks to the
$\Delta$-sum rule.\\
  Some results are also available about the massless flows between
  N=1 unitary
  superconformal models, thanks to perturbative RG analysis
  \cite{pogossyan}. In particular, the results obtained in this
  article
  indicate that (at least for $l$ large), there exist flows
  $\Phi_{2,2} \to \Phi_{2,2}$ from $\mathcal{M}(l+3,4)$ to
  $\mathcal{M}(l+1,4)$, (the case we are particularly interested in is given by
  $l=2$, which
  is not covered by RG analysis).\\
 Consequently, we find it natural enough~-let us recall that the operator $\Phi_{2,2}$ in the coset
$\mathcal{M}(m,k+2)$, which conformal weight is
  $\Delta_{2,2}=\frac{3k}{4m(m+k)}$, is the
fundamental Landau-Ginzburg field-~ to conjecture that we are
dealing with $\Phi \equiv \Phi_{2,2}$ in $\mathcal{M}(k+3,k+2)$,
flowing to $\Phi_{2,2}$ in $\mathcal{M}(3,k+2)$. The conformal
dimension of the UV operator
  $\Phi_{2,2}$ is:
   $$
   \Delta_{2,2}^{\textrm{{\tiny UV}}}=\frac{3k}{4(k+3)(2k+3)},
   $$
   and we conjecture that the variation of the conformal dimension along the flow is
   given by:
\begin{eqnarray}
D_k^{\textrm{\tiny exact}}=\Delta_{2,2}^{\textrm{{\tiny UV}}}
-\Delta_{2,2}^{\textrm{{\tiny IR}}}=\Delta_{2,2}^{\textrm{{\tiny
UV}}} -\Delta_{1,2}^{\textrm{{\tiny
IR}}}=-\frac{k^2}{2(k+3)(2k+3)}. \nonumber
\end{eqnarray}
In Table \ref{tab1} we present the numerical estimation versus the
exact result: we find a good agreement between them, at least
within the precision of the four-particle approximation (the
discrepancies given in the last column of Table 1 are compatible
with such an approximation). Let us note also that the accuracy of
the results observed here is actually better than the one we had
obtained for the $c$-theorem in \cite{GP}, where it had appeared
that the precision of the numerical results diminished as one
increased $k$. In our previous paper, we had linked this
phenomenon to the fact that the conformal weight of the trace
operator
\begin{eqnarray}
\Delta^{\Theta} = \bar{\Delta}^{\Theta}=1-\frac{2}{2 k + 3},
\nonumber
\end{eqnarray}
gets closer to one as $k$ increases, spoiling thus the convergence
in the UV of the integral expressing the variation of the central
charge.

 The funny pattern observed in Table 1 is a bit more
difficult to explain; we believe it could be caused by a combined
effect of $\Delta^{\Theta}\to 1$ when $k\to \infty$, as well as
the non monotonic behaviour of $\Delta^{\textrm{{\tiny
UV}}}_{2,2}$, as $\Delta^{\textrm{{\tiny UV}}}_{2,2}\to 0$ for
both $k\to 0$ and $k\to \infty$.

In any case, we have little doubt that the $\Delta$-sum rule
supports our conjecture. Still, it would be interesting to confirm
our result by means of other methods (we have in mind the TBA
analysis, in a similar way to what has been done in \cite{KM}).
\begin{center}
\begin{table}
\begin{center}
\begin{tabular}{|c|l|l|r|l|l|}
\hline
$k$ &  $D_k^{\textrm{\tiny num}}$ & $D_k^{\textrm{\tiny exact}}$ & \% dev. &
$\Delta_{2,2}^{\textrm{{\tiny UV}}}$ & $\Delta^{\Theta} $ \\
\hline \hline
$1$ & $-0.0254(2)$ & $-0.025$ & $1.6 \% $ & $0.0375$ & $0.6$\\
$2$ & $-0.0609(3)$ & $-0.057142 \dots$ & $6.6 \% $ & $0.042857 \dots$ & $ 0.714286 \dots$\\
$3$ & $-0.0910(3)$ & $-0.083333 \dots$ & $9.2 \% $ & $0.041666 \dots$ & $0.777777 \dots$ \\
$4$ & $-0.1152(5)$ & $-0.103896 \dots$ &$10.9 \% $ & $0.038961\dots$ & $0.818181 \dots$\\
$6$ & $-0.151(2)$ & $-0.133333 \dots$ & $13.2 \% $ & $0.033333 \dots$ & $0.866666 \dots$\\
$10$ & $-0.189(1)$ & $-0.167224 \dots$ & $13.0 \%$ & $0.025083\dots$ & $0.913043 \dots$\\
$20$ & $-0.220(1)$ & $-0.202224 \dots$ & $8.8 \%$ & $0.015166 \dots$ & $0.953488 \dots$ \\
$50$ & $-0.242(2)$ & $-0.228979 \dots$ & $5.7 \%$ & $0.006869\dots$ & $0.980583 \dots$\\
$100$ & $-0.246(2)$ & $-0.239131 \dots$ & $3.0 \%$ & $0.003586 \dots$& $0.990148 \dots$\\
\hline \hline
$\infty$ & $-0.251(1)$ & $-0.25 $ & $0.8 \%$ & $0$ & $1$ \\
\hline
\end{tabular}
\end{center}
\caption{Four particle approximation~-~comparison between
$D_k^{\textrm{\tiny num}}$ and $D_k^{\textrm{\tiny exact}}$
(conjectured variation of the conformal dimension).} \label{tab1}
\end{table}
\end{center}

\section{Form factors of an operator $\Psi$ that flows to the chiral
components
 of $\Phi_{2,1}$ in the $M_{p}$ model.}
\subsection{Energy operator in the flow between TIM and
IM.}
We recall that for the massless flow between TIM and IM, the
asymptotic states consist of right and
 left Majorana fermions, with Lorentz spin $\mathrm{s}=\pm 1/2$.
The form factors
  of the energy operator $\epsilon$ have non zero matrix elements for an odd number of right particles
   and an odd number of left particles.
  They satisfy the residue equation in the $RR$ channel at
$\theta_1=\theta_{2r+1}+\mathrm{i}\pi$ (we recall that
$S_{RR}=S_{LL}=-1$):
\begin{eqnarray}
\lefteqn{\mathrm{res}F^{\epsilon}_{2r+1,2l+1}(\theta_1,\cdots,\theta_{2r+1};\theta'_1,\cdots,\theta'_{2l+1})} \nonumber\\
&&
=-2\mathrm{i}\;F^{\epsilon}_{2r-1,2l+1}(\theta_2,\cdots,\theta_{2r};\theta'_1,\cdots,\theta'_{2l+1})
\left(1-(-1)^{2r-1}\prod_{k=1}^{2l+1}S_{RL}(\theta_{2r+1}-\theta'_k)\right)\nonumber
\end{eqnarray}
and a similar relation in the $LL$ channel. It was noticed in
\cite{P} that they could be written as:
\begin{eqnarray}
F_{2r+1,2l+1}^{\epsilon}(\theta_1,\ldots,\theta_{2r+1};\theta'_1,\ldots,\theta'_{2l+1})=
\frac{\mathrm{i}M}{\mathcal{K}}\prod_{1\leq i<j\leq
2r+1}\sinh\frac{\theta_{ij}}{2} \prod_{1\leq i<j\leq
2l+1}\sinh\frac{\theta'_{ij}}{2}
\nonumber \\
\times \;\prod_{i,j}f_{RL}(\theta_i -\theta'_j)\;
Q_{2r+1,2l+1}^{\epsilon}(\theta_1,\ldots,\theta_{2r+1};\theta'_1,\ldots,\theta'_{2l+1}).
\nonumber
\end{eqnarray}
The expression for $Q_{2r+1,2l+1}^{\epsilon}$ being given by:
\begin{eqnarray}
  \lefteqn{Q^{\epsilon}_{2r+1,2l+1}(\theta_1,\ldots,
    \theta_{2r+1};\theta'_1,\ldots,\theta'_{2l+1})=}
  \nonumber \\
  &&\pi^{r+l}\sum_{T \subset S, \atop \#T=r}\sum_{T' \subset S',\atop
    \#T'=l}\prod_{i \in T, \atop k\in
    \bar{T}}\phi(\theta_{ik})\prod_{i \in T', \atop k\in
    \bar{T}'}\phi(\theta'_{ik}) \prod_{i \in T, \atop k\in
    \bar{T}'}\Phi(\theta_i-\theta'_k) \prod_{i \in T', \atop k\in
    \bar{T}}\tilde{\Phi}(\theta_k-\theta'_i)\nonumber,
\end{eqnarray}
where the functions $\phi(\theta)$ and $\Phi(\theta)$ are defined
in (\ref{phi}) and (\ref{grandphi}) above, and the first recursion
step is \cite{DMS}:
\begin{eqnarray}
F_{1,1}^{\epsilon}(\theta_1;\theta'_1)=\frac{\mathrm{i}M}{\mathcal{K}}f_{RL}(\theta_1
-\theta'_1). \nonumber
\end{eqnarray}
In the IR, $F_{1,1}^{\epsilon}(\theta_1;\theta'_1)\to
Me^{\frac{1}{2}(\theta-\theta')} \sim \psi\bar{\psi}\sim
\epsilon_{\mathrm{Ising}}$, with conformal dimension
 $\Delta_{\textrm{\tiny IR}}=\bar{\Delta}_{\textrm{\tiny IR}}=1/2$. The authors of \cite{DMS}
 checked numerically that the power law behaviour in the UV of the
 two point correlation function
 truncated to one right and one left particle agreed with the
 expected conformal dimension $(1/10,1/10)$ of the field
 $\Phi_{1,2}$ in TIM.

\subsection{Generalizations}
We generalize the previous results for any value of $p$: the
important observation is that the primary field $\Phi_{2,1}$ in
$M_{p}$ has conformal dimension
$\Delta_{2,1}=\bar{\Delta}_{2,1}=\frac{1}{4}+\frac{3}{4p}$: for
$p=3$ it is nothing but the energy operator $\epsilon
=\bar{\psi}\psi$ in the Ising model, and for $p=+\infty$, it
coincides with the spinon field of the $SU(2)_1$ WZNW model. It is
thus natural to think of each chiral components as a generalized
fermion for $p$ arbitrary (in other words, its chiral components
generate the asymptotic states) \cite{smirnov3}.\\
We call $g_{\uparrow\downarrow}$ and
$\bar{g}_{\uparrow\downarrow}$ the holomorphic components of
$\Phi_{2,1}$: $g_{\uparrow}$,$\;\bar{g}_{\uparrow}$ have Lorentz
spin $\Delta_{2,1}$, whereas
$g_{\downarrow}$,$\;\bar{g}_{\downarrow}$ have Lorentz spin
$-\Delta_{2,1}$. These operators are topologically charged: as
$g_{\uparrow}$,$\;\bar{g}_{\downarrow}$ create asymptotic
particles, they have topological charge $+1$. Analogously,
$g_{\downarrow}$,$\;\bar{g}_{\uparrow}$ have topological charge
$-1$. The problem of the construction of form factors of the
"parafermionic" operators $g_{\uparrow},\bar{g}_{\downarrow}$ with
a non zero topological charge in the RSOS restriction of the
Sine-Gordon model was first addressed in \cite{smirnov3}, the form
factors of the components of the Fermi field in Sine-Gordon were
constructed in \cite{BFKZ}, and quite generally, form factors of
topologically charged operators can be found in \cite{LZ}.
\begin{itemize}
\item $(1,1)$ topological charge:\\
We shall look now at form factors of an operator $\Psi$ which
flows to $g_{\uparrow}\bar{g}_{\downarrow}$.
 The topological charge is equal to $1$ in both the right and left sectors.
The first step of the recursion relation is given by:
\begin{eqnarray}
F_{1,1}(\theta_1,\theta'_1)_{++}=\frac{\mathrm{i}M^{2\Delta_{2,1}}}{\mathcal{K}}
f_{RL}(\theta_1-\theta'_1)e^{(\Delta_{2,1}-\frac{1}{2})(\theta_1-\theta'_1)},
\label{firststep}
\end{eqnarray}
such that in the IR limit $\theta_1 -\theta'_1\to -\infty$:
$$
F_{1,1}(\theta_1,\theta'_1)_{++} \to
f_1^{g_{\uparrow}}(\theta_1)\;
f_1^{\bar{g}_{\downarrow}}(\theta'_1).
$$
We make the following ansatz for the solution:
\begin{eqnarray}
\lefteqn{F_{2r+1,2l+1}(\theta_1,\dots,\theta_{2r+1};\theta'_{1},\dots,\theta'_{2l+1})=}\nonumber \\
&& \frac{M^{2\Delta_{2,1}}}{\mathcal{K}}\;\prod_{1\le i<j\le
2r+1}f_{p}(\theta_{ij})\prod_{1\le i<j\le
2l+1}f_{p}(\theta'_{ij})\;
\prod_{i,j}f_{RL}(\theta_i-\theta'_{j})\quad \times
 \nonumber \\
&& \int_{C_{\theta}}\prod_{m=1}^{r} du_m\;
h_{RR}(\theta,u)\;p^{-\Delta_{2,1}+\frac{1}{2p}}_{2r+1}(\theta,u)
 \tilde{\Psi}(\theta,{u})\; \int_{C_{\theta'}}\prod_{m=1}^{l} dv_m\; h_{LL}(\theta',v)\;
p^{\Delta_{2,1}+\frac{1}{2p}}_{2l+1}(\theta',v) \tilde{\Psi}(\theta',{v})\nonumber \\
&& \times \quad
\mathcal{R}_{2r+1,2l+1}(\theta,\theta',{u},{v}).\nonumber
\end{eqnarray}
In the formula above:
\begin{itemize}
\item
we introduced the modified Bethe ansatz states
\begin{eqnarray}
\tilde{\Psi}^{p}_{\epsilon_1 \epsilon_2 \dots
\epsilon_{2n+1}}(\theta,u)\equiv e^{\frac{1}{2p}\sum_i \epsilon_i
\theta_i} \Psi^{p}_{\epsilon_1 \epsilon_2 \dots
\epsilon_{2n+1}}(\theta,u), \quad \epsilon_i=\pm,\;
\sum_{i=1}^{2n+1}\epsilon_i=\pm 1, \nonumber
\end{eqnarray}
One should pay attention that under a Lorentz transformation, this
multiparticule state possesses a Lorentz spin
$\mathrm{s}=\frac{1}{2p}$ when $\sum_{i=1}^{2n+1}\epsilon_i=1$.
\item
the $p$-function for $g_{\uparrow}$ with topological charge $1$
is:
$$
p_{2n+1}^{\mp\Delta_{2,1}+\frac{1}{2p}}(\theta,u)=\frac{1}
{e^{\mp\mathrm{i}\pi\Delta_{2,1}}}
\frac{\prod_{m=1}^{n}e^{\left(\mp
2\Delta_{2,1}+\frac{1}{p}\right)u_m}}{\prod_{i=1}^{2n+1}
e^{\left(\mp\Delta_{2,1}+\frac{1}{2p}\right)\theta_i}}.
$$
Under a Lorentz transformation, this $p$-function possesses a
Lorentz spin $\mathrm{s}=\pm\Delta_{2,1}-1/2p$, such that together
with the Bethe ansatz state, the total Lorentz spin is equal to
$\mathrm{s}=\pm\Delta_{2,1}$.
\item
we took into account the relation between the number of particles
$n$, the topological charge $q$ and the number of integration
variables $m$: $q=n-2m$.
\end{itemize}

 Let the set
$S=(1,\dots,2r+1)$, and $T\subset S$ and $\bar{T}\equiv
S\backslash T$. These subsets have the number of elements:
$\#T=r+1$, $\#\bar{T}=r$,
\begin{eqnarray}
T = \{i_1<i_2<\dots<i_{r+1}\},\quad \bar{T} =
\{k_1<k_2<\dots<k_{r}\}. \nonumber
\end{eqnarray}
The sets $S',T'$ and $\bar{T}'$ are similarly defined. We propose:
\begin{eqnarray}
\lefteqn{\mathcal{R}_{2r+1,2l+1}(\theta,\theta',{u},{v})=\mathrm{i}^{r+l+1}\sum_{T
\subset S,\atop \# T=r+1}\sum_{T' \subset S',\atop \# T'=l+1}
 \frac{\prod_{k,l\in \bar{T}\atop k<l
}\cos\frac{\theta_{kl}}{2\mathrm{i}}} {\prod_{i\in T,\atop k\in
\bar{T}}\sin\frac{\theta_{ik}}{2\mathrm{i}}}e^{\frac{1}{2}\sum\theta_{ki}}
\frac{\prod_{k,l\in \bar{T}'\atop
k<l}\cos\frac{\theta'_{kl}}{2\mathrm{i}}}{\prod_{i\in T',\atop
k\in \bar{T}'}
\sin\frac{\theta'_{ik}}{2\mathrm{i}}}e^{\frac{1}{2}\sum\theta'_{ik}}}
\nonumber
\\
&&\times \;
 \frac{\prod_{i\in T,\atop m=1,\dots
,r}\cos\frac{\theta_i-u_m}{2\mathrm{i}} \prod_{i\in T',\atop
m=1,\dots ,l}\cos\frac{\theta'_i-v_m}{2\mathrm{i}}}{\prod_{1\le
m<n\le r} \cos \frac{u_m-u_n}{2\mathrm{i}}\prod_{1\le m<n\le
l}\cos \frac{v_m-v_n}{2\mathrm{i}}}\prod_{i \in T, \atop k\in
      \bar{T}'}\tilde\Phi(\theta_i-\theta'_k) \prod_{i \in T', \atop k\in
      \bar{T}}\Phi(\theta_k-\theta'_i).
\nonumber
\end{eqnarray}
The function $\mathcal{R}_{2r+1,2l+1}$ satisfies the properties at
$\theta_{1}=\theta_{2r+1}+\mathrm{i}\pi$:
\begin{itemize}
\item
 $u_{r}=\theta_{2r+1}$:
\begin{eqnarray}
\lefteqn{\mathcal{R}_{2r+1,2l+1}(\theta_1,\dots,\theta_{2r};\theta';u_1,\dots
,u_{r};v)=}
\nonumber \\
&& -\mathcal{R}_{2r-1,2l+1} (\theta_2,\dots,\theta_{2r+1};\theta';
u_1,\dots, u_{r-1};v)
 \prod_{k=1}^{2l+1}\tilde\Phi(\theta_{2r+1}-\theta'_k).
\label{q1}
\end{eqnarray}
\item
 $u_{r}=\theta_{2r+1}\pm \mathrm{i}\pi$:
\begin{eqnarray}
\lefteqn{\mathcal{R}_{2r+1,2l+1}(\theta_1 \dots
\theta_{2r+1};\theta';
u_1 \dots u_{r};v)=}\nonumber \\
&& \mathcal{R}_{2r-1,2l+1} (\theta_2,\dots,\theta_{2r};\theta';
u_1,\dots, u_{r-1};v)
\prod_{k=1}^{2l+1}\Phi(\theta_{2r+1}-\theta'_k), \label{q2}
\end{eqnarray}
\end{itemize}
and similar relations in the $LL$-channel. In particular
$\mathcal{R}_{1,1}=\mathrm{i}e^{\frac{1}{2}(\theta'_1-\theta_1)}$.
We have in the IR (checked for a small number of particles with
Mathematica):
$$
F_{2r+1,2l+1}(\theta_1,\cdots,\theta_{2r+1};\theta'_1,\cdots,\theta'_{2l+1})
\to f_{2r+1}^{g_{\uparrow}}(\theta_1,\cdots,\theta_{2r+1})\;
f_{2l+1}^{\bar{g}_{\downarrow}}(\theta'_1,\cdots,\theta'_{2l+1}).
$$
Let us note that the minus sign in the equation (\ref{q1}) is
related to the fact that the term $\prod_{k=1}^{2l+1}
S_{RL}(\theta_{2r+1}-\theta'_k)$ in the residue equation gives an
extra minus sign in the IR limit.
\item $(-1,-1)$ topological charge\\
This case is similar to the previous one, at the condition of
changing the number of integration variables $(r,l)$ into
$(r+1,l+1)$, introducing a new function $\mathcal{P}$:
\begin{eqnarray}
\mathcal{P}_{2r+1,2l+1}(\theta,\theta',{u},{v})=
(-\mathrm{i})^{r+l-1}\sum_{T \subset S, \atop \#T=r}\sum_{T'
\subset S',\atop \#T'=l} \frac{\prod_{k,l\in \bar{T}\atop k<l
}\cos\frac{\theta_{kl}}{2\mathrm{i}}} {\prod_{i\in T,\atop k\in
\bar{T}}\sin\frac{\theta_{ik}}{2\mathrm{i}}}e^{\frac{1}{2}\sum\theta_{ik}}
\frac{\prod_{k,l\in \bar{T}'\atop
k<l}\cos\frac{\theta'_{kl}}{2\mathrm{i}}}{\prod_{i\in T',\atop
k\in \bar{T}'}
\sin\frac{\theta'_{ik}}{2\mathrm{i}}}e^{\frac{1}{2}\sum\theta'_{ki}}\nonumber
\\\times \;
 \frac{\prod_{i\in T,\atop m=1,\dots
,r+1}\cos\frac{\theta_i-u_m}{2\mathrm{i}} \prod_{i\in T',\atop
m=1,\dots ,l+1}\cos\frac{\theta'_i-v_m}{2\mathrm{i}}}{\prod_{1\le
m<n\le r+1} \cos \frac{u_m-u_n}{2\mathrm{i}}\prod_{1\le m<n\le
l+1}\cos \frac{v_m-v_n}{2\mathrm{i}}}\prod_{i \in T, \atop k\in
      \bar{T}'}\tilde\Phi(\theta_i-\theta'_k) \prod_{i \in T', \atop k\in
      \bar{T}}\Phi(\theta_k-\theta'_i),
\nonumber
\end{eqnarray}
that satisfies similar equations to (\ref{q1}) and (\ref{q2}).
There are some obvious modifications in the $p$-functions to be
made, that we think are needless to make more explicit here.
\end{itemize}

\subsection{Numerical results}
In the flow TIM $\to$ IM, the UV operator $\Phi_{1,2}$ with
conformal dimension $(1/10,1/10)$ flows in the IR to $\Phi_{2,1}$
with
conformal dimension $(1/2,1/2)$.\\
Let us notice the operators $\Phi_{2,1}$ in
$M_{k+2}=\mathcal{M}(k+2,3)$ with conformal dimension
$\Delta_{2,1}^{\textrm{\tiny IR}}=\frac{1}{4}+\frac{3}{4(k+2)}$
coincide with the operators $\sigma_1\Phi_{2,1}$ in the coset
$\mathcal{M}(3,k+2)$ (for $k=1$, it is $\Phi_{2,1}$).\\
Consequently, by analogy with the case $k=1$, we are tempted to
conjecture that the UV operator $\Psi$ we are looking for is
$\sigma_1\Phi_{1,2}$ in the coset model $\mathcal{M}(k+3,k+2)$.
This operator has conformal dimension\footnote{More generally, the
operator $\sigma_1\Phi_{2,1}$ in the coset model
$\mathcal{M}(k+2,l+2)$ has conformal dimension
$\Delta=\Delta_{2,1}^{M_{k+2}}+\Delta_{2,1}^{M_{l+2}}-\frac{1}{2}$,
where $\Delta_{2,1}^{M_{k+2}}$ ($\Delta_{2,1}^{M_{l+2}}$) is the
conformal dimension of $\Phi_{2,1}$ in the minimal model $M_{k+2}$
($M_{l+2}$). This operator is the generalization of $\Phi_{2,1}$
in minimal models.}:
\begin{eqnarray}
\label{psiUV}
\Delta^{\textrm{{\tiny UV}}}=\frac{3(k+1)}{4(k+2)(2k+3)}.
\end{eqnarray}
In the framework of form factors we can try to give some
evidence in favour of
such a conjecture by means of the analysis of the $\langle \Psi(x)
\Psi(0)\rangle$ correlation function. As usual, we will write down the
leading contribution to the spectral expansion for the correlator given by
the two-particles form factor (\ref{firststep})
\begin{eqnarray}
\langle \Psi(x) \Psi(0)\rangle \sim \int_{-\infty}^{\infty} d
\theta \; |F_{1,1}(\theta)_{++}|^2\; K_0 (M r \,e^{\theta/2}),
\nonumber
\end{eqnarray}
where $K_0(z)$ is the modified Bessel function of order zero.

The next step is then to compare the approximated
correlator with the expected power-law behaviour in both the IR and UV,
 being $\sim r^{-4 \Delta^{\textrm{{\tiny IR}}}_{2,1}}$ and
 $\sim r^{-4 \Delta^{\textrm{{\tiny UV}}}}$ respectively.
In other words, we are interested in comparing the slope of the
approximated correlator with that predicted by CFT at the critical
points (we will use the log-log plane to plot them).

\begin{figure}

\centerline{\psfig{figure=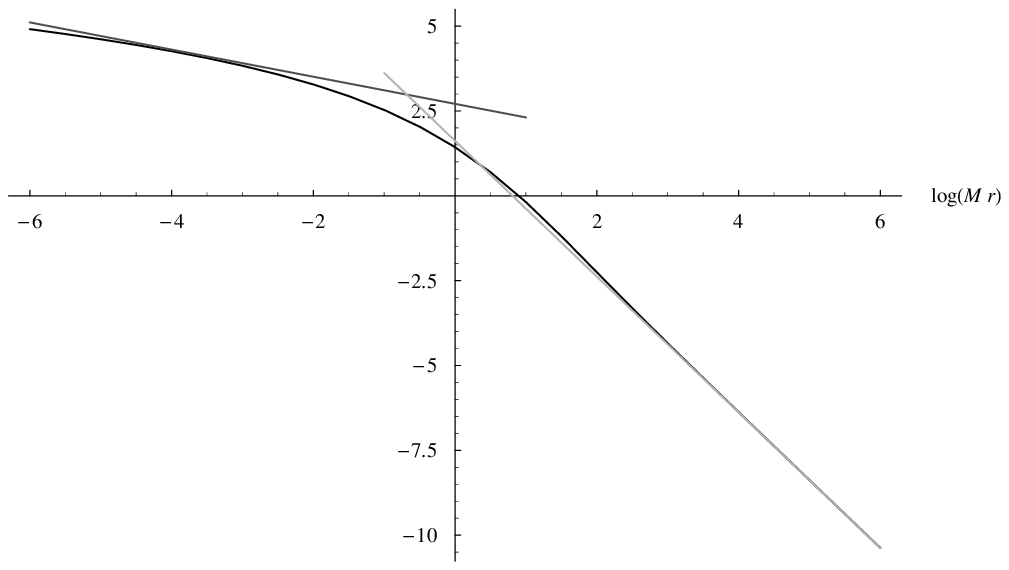,width=0.7\textwidth}}
\caption{Flow $k=1$, logarithmic plot of the correlator
$\langle \Psi(x) \Psi(0) \rangle$ (black line) together
with both the IR (grey line) and the UV (dark-grey line)
behaviours.} \label{kp_1}
\end{figure}

\begin{figure}
\centerline{\psfig{figure=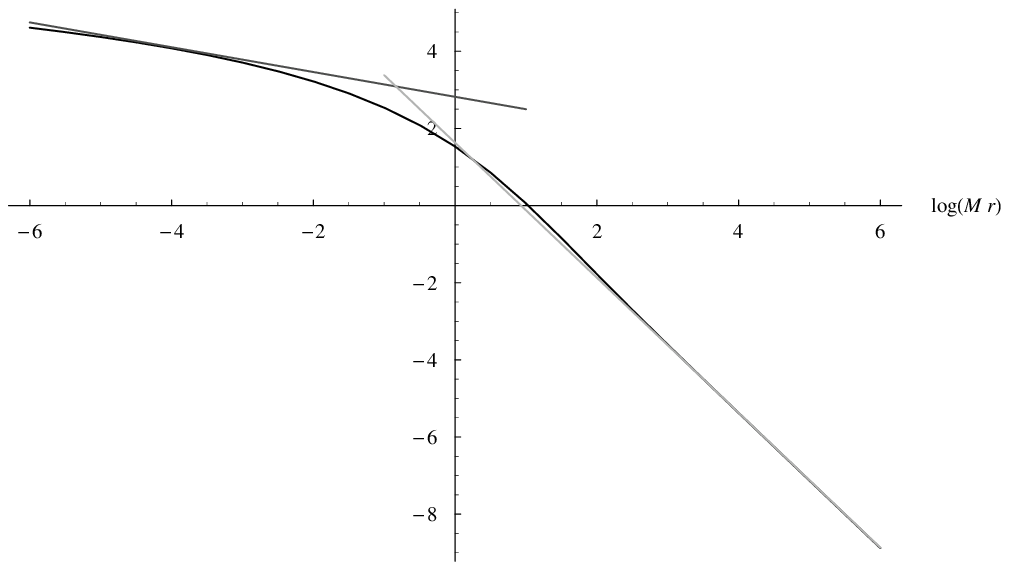,width=0.7\textwidth}}
\caption{Flow $k=2$, logarithmic plot of the correlator
$\langle \Psi(x) \Psi(0) \rangle$ (black line) together
with both the IR (grey line) and the UV (dark-grey line)
behaviours.} \label{kp_2}
\end{figure}

\begin{figure}
\centerline{\psfig{figure=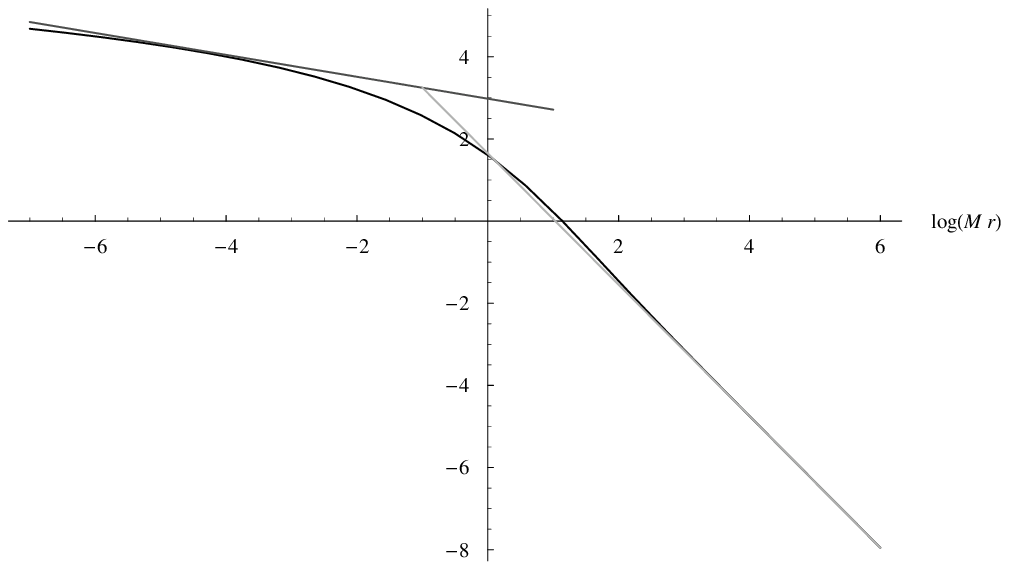,width=0.7\textwidth}}
\caption{Flow $k=3$, logarithmic plot of the correlator
$\langle \Psi(x) \Psi(0) \rangle$ (black line) together
with both the IR (grey line) and the UV (dark-grey line)
behaviours.} \label{kp_3}
\end{figure}

\begin{figure}
\centerline{\psfig{figure=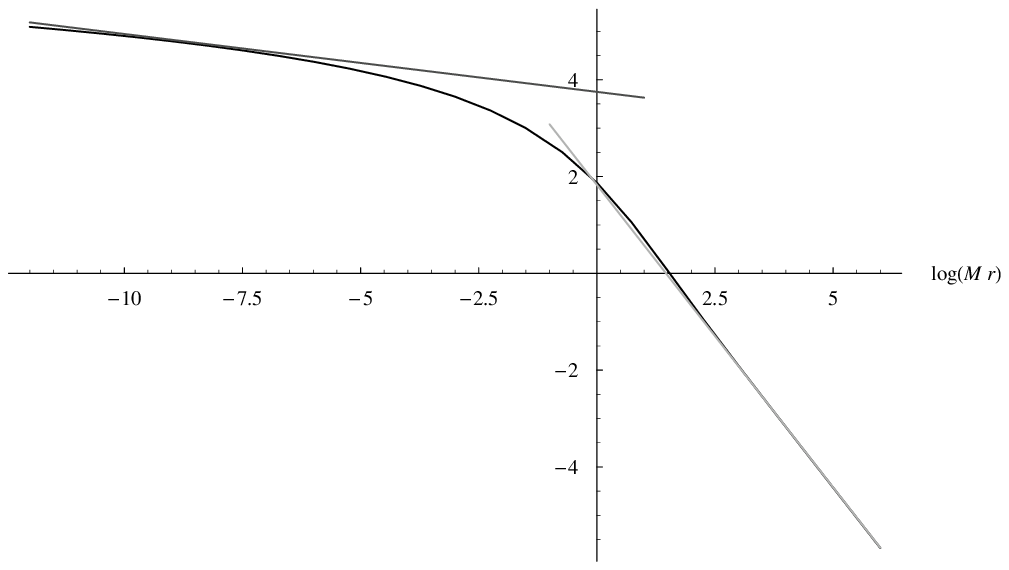,width=0.7\textwidth}}
\caption{Flow $k=10$, logarithmic plot of the correlator
$\langle \Psi(x) \Psi(0) \rangle$ (black line) together
with both the IR (grey line) and the UV (dark-grey line)
behaviours.} \label{kp_10}
\end{figure}

The inspection of the diagrams (see figures \ref{kp_1},
\ref{kp_2}, \ref{kp_3}, \ref{kp_10}, obtained for $k=1$, $2$, $3$,
$10$ respectively) shows first of all the expected agreement in
the IR limit: it is worth recalling that as in the case of the
trace operator \cite{GP}, the leading contribution of the spectral
series is enough to obtain the exact IR power-law.

It is instead quite surprising that such an approximation is able to give
a qualitative good agreement also in the UV (again similar to the case of
the stress-energy tensor) when (\ref{psiUV}) is conjectured to give the correct
power-law behaviour.

On the one hand, the previous result can be considered as a
qualitative evidence that the operator $\sigma_1\Phi_{1,2}$ in the
UV actually flows to the operator $\Phi_{2,1}$ in the IR theory.
On the other hand, we would like to stress that such a result,
being qualitative, is far from being neither conclusive nor
satisfactory in the perspective of the identification of the UV
operator $\sigma_1\Phi_{1,2}$. At best it can be viewed as an
indication to stimulate the research of other, more reliable,
methods to face the problem of the identification of operators in
massless flows.

A final remark about the diagrams: since we are interested in the
comparison of the slope of the curves, all the normalizations have
been fixed in order to make such a comparison as clear as
possible.

\section*{Concluding remarks}
In this work, we constructed form factors of operators in the
massless flow from the coset model $su(2)_{k+1} \otimes su(2)_k
/su(2)_{2k+1}$ to the $ M_{k+2} $ minimal model, mimicking the
construction done in \cite{DMS} of form factors of the
magnetization operator and the energy operator in the massless
flow \\TIM $\to$ IM.

What we did in the section 1.2 is to look for solutions of the
residue equation that are obtained replacing the $RR$ and $LL$
$S$-matrices $S^{RSOS}_3=-1$ by the $S^{RSOS}_{k+2}$-matrix, given
that they should reproduce in the IR limit in both the right and
left channel the form factors of the operator $\Phi_{1,2}$ in the
minimal model $M_{k+2}$. Then we made a numerical check on the
variation of conformal weight along the flow thanks to the
$\Delta$-sum rule, and found that, if not excellent, it is
compatible with
 the hypothesis that we are dealing with the operator $\Phi_{2,2}$ in the UV.

The situation is slightly more complicated in the section 2.2,
because the asymptotic particles possesses generalized statistics
for $k\neq 1$ : we took into account the IR properties only, and
constructed form factors which in the IR limit reproduce the form
factors of the parafermionic operators with conformal dimension
$(\pm \Delta_{2,1},0)$ and $(0,\pm \Delta_{2,1})$. Notwithstanding
this, the approximation of the two point function with the lowest
form factor with one right and one left particle is enough to
give, at least at a qualitative level, a good agreement with the
power-law behaviour expected in the UV if we conjecture that the
corresponding operator is $\sigma_1 \Phi_{1,2}$.

We are not saying that we constructed all the possible flows of
operators, but only those which have an obvious counterpart in the
flow TIM $\to$ IM. It would be interesting to know what else could
be constructed.

The results obtained in both \cite{GP} and the present work show
that form factors in integrable massless models can provide
important non perturbative information, even in more complicated
cases than the flow TIM~$\to$~IM. Obviously, we are in a
privileged situation: having a one parameter family of flows at
hand certainly allows us to understand better the loss of
precision in the numerical tests for the $c$- and $\Delta$-sum
rules as one increases the parameter $k$. Had we worked on the
flow $\textrm{PCM}_1 \to SU(2)_1$ only, the discrepancy of $43\%$
with respect to the exact value of the central charge \cite{GP}
would have probably led us to conclude that our 4-particle form
factor for the trace operator was wrong! Interestingly enough, the
results of the present work as well as those of \cite{P2,GP} show
that whether in the massive or the massless case, the truncation
to the lowest form factor does not systematically give a 'very
accurate' approximation of the correlation function. One can
really wonder up to what point it can be unaccurate; certainly,
this means that one has to be rather cautious when interpreting
the numerical results, in the case where a large discrepancy with
respect to the expectations is observed.

We are aware of the fact that our task was considerably simplified
as the flow is along $T\bar{T}$. For other integrable massless
flows with a non diagonal $RL$ scattering, the situation is far
more involved, both theoretically and numerically: in most cases,
we do not expect that the lowest form factors can be nicely
written as an explicit product of simple functions in the right
and left channel as it is the case here; likely, even with the
lowest number of particles one might not get rid of the
integration variables, thus the integrals for the form factors
should be evaluated numerically first.

\section*{Acknowledgments}
The work of P.G. was supported by the Euclid Network
HPRN-CT-2002-00325. B.P. was supported by a Linkage International
Fellowship of
 the Australian Research Council.

\newpage


\begin{thebibliography}{99}

\bibitem{GP} P.~Grinza and B.~Ponsot, "Form factors in the massless coset
models $su(2)_{k+1} \otimes su(2)_k /su(2)_{2k+1}$ -~Part I",
[arXiv:hep-th/0411043], to appear in Nucl.~Phys.~B.

\bibitem{CSS} C.~Crnkovic, G.M.~Sotkov and M.~Stanishkov, Phys.~Lett.~{\bf B226} (1989) 297.

\bibitem{GKO}
P.~Goddard, A.~Kent and D.I.~Olive, Phys.~ Lett.~{\bf B152} (1985)
88.

\bibitem{DMS}
G.~Delfino, G.~Mussardo and P.~Simonetti, Phys.~Rev.~{\bf D51}
(1995) 6620 [arXiv:hep-th/9410117].


\bibitem{KW} M.~Karowski and P.~Weisz, Nucl.~Phys.~\textbf{B139}
(1978) 455.


\bibitem{BKW} B.~Berg, M.~Karowski and P.~Weisz,
  Phys.~Rev.~\textbf{D19} (1979) 2477


\bibitem{smirnov2}
F.A.~Smirnov, {\it "Form factors in completely integrable models
of Quantum Field theory"}, Adv.~Series in Math.~Phys.~\textbf{14},
World Scientific 1992.

\bibitem{Z2} Al.B.~Zamolodchikov, Nucl.~Phys.~\textbf{B358} (1991)
 524.

\bibitem{BFKZ} H.~Babujian, A.~Fring, M.~Karowski and A.~Zapletal, Nucl.~Phys.~\textbf{B538}
 (1999) 535 [arXiv:hep-th/9805185].


\bibitem{BK} H.~Babujian and M.~Karowski, Nucl.~Phys.~\textbf{B620}
(2002) 407 [arXiv:hep-th/0105178].



\bibitem{Bernard}
D.~Bernard, Phys.~ Lett.~ {\bf B279} (1992) 78
[arXiv:hep-th/9201006].



\bibitem{L-RS}
A.~LeClair, Phys.~ Lett.~{\bf B230} (1989) 103, D.~Bernard and
A.~Leclair, Nucl.~Phys.~{\bf B340} (1990) 721, F.A.~Smirnov,
Int.~J.~Mod.~Phys.~{\bf A4} (1989) 4213 and Nucl.~Phys.~{\bf B337}
(1990) 156

\bibitem{RS}
 N.Yu.~Reshetikhin and F.A.~Smirnov, Commun.~Math.~ Phys.~{\bf 131} (1990) 157.




\bibitem{ZZ} A.B.~ Zamolodchikov and Al.B.~ Zamolodchikov, Nucl.~Phys.~\textbf{B379} (1992)
 602.


\bibitem{YZ} V.P.~Yurov and Al.B.~Zamolodchikov, Int.~J.~Mod.~Phys.~\textbf{A6} (1991)
3419.

\bibitem{P}
B.~Ponsot, Phys.~Lett.~{\bf B575} (2003) 131
[arXiv:hep-th/0304240].




\bibitem{CM} J.L.~Cardy and G.~Mussardo, Nucl.~Phys.~\textbf{B340} (1990)
387.



\bibitem{L}
S.L.~Lukyanov, Mod.~Phys.~Lett.~{\bf A12} (1997) 2543
[arXiv:hep-th/9703190].


\bibitem{P2}
B.~Ponsot, ``Form factors in the $SS$ model and its RSOS
restrictions'', [arXiv:hep-th/0405218].

\bibitem{DSC}
G.~Delfino, P.~Simonetti and J.L.~Cardy,
Phys.~Lett.~{\bf B387} (1996) 327 [arXiv:hep-th/9607046].


\bibitem{Z4}
A.B.~Zamolodchikov, Sov.~J.~Nucl.~Phys.~{\bf 46} (1987) 1090
[Yad.~Fiz.~{\bf 46} (1987) 1819].

\bibitem{LC}
A.W.W.~Ludwig and J.L.~Cardy, Nucl.~Phys.~{\bf B285} (1987) 687.

\bibitem{Z5}
A.B.~Zamolodchikov, Sov.~J.~Nucl.~Phys.~{\bf 44} (1986) 529
[Yad.~Fiz.~{\bf 44} (1986) 821].

\bibitem{KM}
T.R.~Klassen and E.~Melzer, Nucl.~Phys.~{\bf B370} (1992) 511.


\bibitem{pogossyan}
R.~Pogossyan, Sov.~J.~Nucl.~Phys.~{\bf 48} (1988) 763.

\bibitem{smirnov3}
F.A.~Smirnov,
Commun.~Math.~Phys.~{\bf 132} (1990) 415.




\bibitem{LZ}
S.L.~Lukyanov and A.B.~Zamolodchikov, Nucl.~Phys.~{\bf B607}
(2001) 437 [arXiv:hep-th/0102079].











\end{thebibliography}
\end{document}